\documentclass[11pt]{article}

\usepackage{troffms}

\begin{document}

\footer{\rm\thepage}

\title{Fokker - Planck equation for incompressible fluid \\
}
\author{Igor A. Tanski\\\
        Moscow, Russia\\
        tanski.igor.arxiv@gmail.com\\
        } 
        
\rule{6in}{1pt}
        
\begin{abstract}
In this article we derive Fokker - Planck equation 
for incompressible fluid and investigate its properties
\end{abstract}

\shead{Keywords}

\par\noindent
Fokker-Planck equation, continuum mechanics, incompressible fluid

\rule{6in}{1pt}

\section{Introduction}

\par
The object of our considerations is a special case of Fokker -
Planck equation, which describes evolution of 3D continuum of
non-interacting particles imbedded in a dense
medium without outer forces.
The interaction between particles and medium combines diffusion in physical space and velocities space.
Classic Fokker - Planck equation contains only two forces, which act on particles: damping force proportional to velocity and random force. 

\par
The aim of this work is to derive equation for the case of constrained movement - the space density of particles remains constant. As a result, our equation contains the third force - constraint reaction. This is equivalent to hydrostatic pressure.

\par
Our starting point is the classical Fokker - Planck equation:

$$
{\partial n  \over \partial t} +
v_k {\partial n  \over \partial x_k} 
- \alpha\  {\partial \over \partial v_j} (v_j n)
= k\  {\partial^2 n  \over \partial v_j \partial v_j} .
\eqno (1)$$

\par\noindent
where
\par\noindent
$n = n(t, x_1 , x_2 , x_3 , v_1 , v_2 , v_3 )$ - density;
\par\noindent
$t$ - time variable;
\par\noindent
$x_1 , x_2 , x_3 $ - space coordinates;
\par\noindent
$v_1 , v_2 , v_3$ - velocities;
\par\noindent
$\alpha$ - coefficient of damping;
\par\noindent
$k$ - coefficient of diffusion.

\par
In the following section we add to this equation terms due to reaction of constraints and thus derive the Fokker - Planck equation for incompressible fluid. Then we rewrite the equation for curvilinear coordinates. In the fourth section we obtain a stationary solution of Fokker - Planck equation for incompressible fluid and in the fifth section we derive linearized equation. In the last section we introduce symmetries of Fokker - Planck equation for incompressible fluid.

\section{Derivation of Fokker - Planck equation for incompressible fluid}

\par
In our previous work (see ref. [1-2]) we deducted equation (1) from the kinematic equation (see below). The crucial point is expression for the force, which acts upon particles.
We represented the force as a sum of random force $R_i$
and friction force, which is proportional to the velocity:

$$
Q_i = R_i - \alpha v_i .
\eqno (2)$$

\par
We calculated random force $R_i$ from the law 

$$
n R_i = - k {\partial n  \over \partial v^i } .
\eqno (3)$$

\par
(2-3) give explicit expressions for the forces. As we show in [2], (2-3) imply equation (1). 

\par
Now we try to get expression for reaction of constraint. We need some additional principle for this purpose. In the following we accept the Gauss principle of least constraint (see [3]). 

\par
Gauss principle reads for discrete system of particles:

$$
G = {1 \over 2 }\sum m_r \left\{
\left(
\dotdot x_r - {X_r  \over m_r }\right)^2 +
\left(
\dotdot y_r - {Y_r  \over m_r }\right)^2 +
\left(
\dotdot z_r - {Z_r  \over m_r }\right)^2 
\right\}
\eqno (4)$$

\par\noindent
where 
\par\noindent
$x_r , y_r , z_r $ - material point coordinates ;
\par\noindent
$\dotdot x_r , \dotdot y_r , \dotdot z_r $ - accelerations ;
\par\noindent
$X_r , Y_r , Z_r $ - components of outer forces, acting 
on this point. They are dependent on coordinates $x_r , y_r , z_r $
and velocities $\dot x_r , \dot y_r , \dot z_r $.

\par
Besides $X_r , Y_r , Z_r $ reactions of constraints are also to be considered, though they are not given explicitly. They appear as Lagrange multipliers of constrains.

\par
Let us suppose, that coordinates $x_r , y_r , z_r $ and velocities $\dot x_r , \dot y_r , \dot z_r $ are kinematically possible, that is they satisfy constraints (holonomic or nonholonomic). Then effective accelerations $\dotdot x_r , \dotdot y_r , \dotdot z_r $ provide minimum of $G$, comparing with all kinematically possible accelerations.

\par
The derivation of reactions of constrains for the simple case of discrete system of particles is straightforward (see [3]). We deal with more complicated case of continuum with dispersed velocities - many particles with different velocities are simultaneously placed in each point of space. 

\par
But first of all we examine first intermediate case of continuum with uniquely defined velocities field, which is simpler. For this case the Gauss principle was used in [4], [5].

\par
For such a continuum (fluid) we have the following expression for accelerations:

$$
a_i = 
{\partial u_i  \over \partial t} +
u_k {\partial u_i  \over \partial x_k}.
\eqno (5)$$

\par
In this way we get the following expression for Gauss functional:

$$
G = \int
{\rho \over 2 }\left(
a_m - {1 \over \rho }F_m
\right)\ 
\left(
a_m - {1 \over \rho }F_m
\right)\ 
dV .
\eqno (6)$$

\par
Condition of incompressibility of the velocities field $u_i$ is 

$$
{\partial u_k  \over \partial x_k} = 0.
\eqno (7)$$

\par
Differentiating (7) on time, we get expression for constraint on accelerations field:

$$
{\partial^2 u_k  \over \partial x_k \partial t} = 0.
\eqno (8)$$

\par
This implies the following constraint on accelerations $a^i$:

$$
{\partial a_k  \over \partial x_k} =
{\partial \dot u_k  \over \partial x_k} +
{\partial  \over \partial x_m}
\left(
u_k {\partial u_m  \over \partial x_k}
\right) = 
{\partial u_k  \over \partial x_m}
{\partial u_m  \over \partial x_k}
 .
\eqno (9)$$

\par
To take in account (9) we use Lagrange multipliers method. This gives following modified expression for Gauss functional:

$$
G = \int
\left[
{\rho \over 2 }\left(
a_m - {1 \over \rho }F_m
\right)\ 
\left(
a_m - {1 \over \rho }F_m
\right) -
p\  
\left(
{\partial a_k  \over \partial x_k} -
{\partial u_k  \over \partial x_m}
{\partial u_m  \over \partial x_k}
\right)
\right]\ 
dV.
\eqno (10)$$

\par\noindent
where $p$ is Lagrange multiplier for constraint (9) . We seek the extreme of functional (10) with $a_j$ as unknown variables.

\par
Variational Euler equations for functional (10) are

$$
\rho\ 
\left(
\dot u_m +
u_i {\partial u_m  \over \partial x_i} 
- {1 \over \rho }F_m
\right)
 +
{\partial p  \over \partial x_m} = 0.
\eqno (11)$$

\par
They are identical to standard Euler equations for incompressible fluid. We see, that Lagrange's multiplier $p$ is pressure and $ 1 / \rho\  ( {\partial p} / {\partial x^k} )$ is expression for reaction of incompressibility constraint.

\par
We conclude, that as for get Fokker - Planck equation for incompressible fluid we should:

\par\noindent
- to add to system additional equation - expression for incompressibility constraint;

\par\noindent
- to add to all forces, acting on particle (see (2)), gradient of the pressure as reaction of ideal constraint .

$$
Q_i = 
R_i - 
{1 \over \rho }{\partial p  \over \partial q^i} -
\alpha v_i .
\eqno (12)$$

\par
In this way we get the following system with one integral constraint expression and one differential equation.

\boxit{
$$
\rho =
\int_V n\  dv_1 dv_2 dv_3 = const ,
\eqno (13)$$
}

\par\noindent
where $\rho$ is constant density. 

\boxit{
$$
{\partial n  \over \partial t} +
v_k {\partial n  \over \partial x_k} - 
\alpha\  { \partial (v_j n)  \over \partial v_j} -
{1 \over \rho }{\partial n  \over \partial v_j}
{\partial p  \over \partial x_j} 
= k\  {\partial^2 n  \over \partial v_j \partial v_j} .
\eqno (14)$$
}

\par
(13) imply equations $({\partial \rho} / {\partial {x_i})} = 0$.

\par
This derivation method is insufficient, because we obtained expression (12) from the wrong model. Now we return to the case of continuum with dispersed velocities - there exist many particles with different velocities in each point of space.

\par
Kinematic of single particle is described by differential equations:

$$
{d x_k  \over dt }= v_k ;
\eqno (15)$$

$$
{d v_k  \over dt }= b_k ;
\eqno (16)$$

\par\noindent
where 
\par\noindent
$x_k$ - space coordinates;
\par\noindent
$v_k$ - velocities of particles;
\par\noindent
$b_k$ - accelerations of particles.

\par
Evolution of particles density $n$ satisfies the following equation:

$$
{\partial n  \over \partial t} +
{\partial ( n v_k )  \over \partial x_k} +
{\partial ( n b_k )  \over \partial v_k}
= 0 .
\eqno (17)$$

\par
If accelerations are independent from $n$,
(15) and (16) are equations of characteristics of (17).

\par
The dynamics of the system is equation of movement of particles - the second Newton's law:

$$
b_j = {1 \over m }Q_j ;
\eqno (18)$$

\par\noindent
where $m$ - particles mass, the force $Q_j$ is the same as in (2-3), (12).

\par
We accept (13) as  constraint on $n$ and try to get expression for constraints on accelerations $b_j$. (13) implies

$$
\int_V 
{\partial n  \over \partial t}
dv_1 dv_2 dv_3 = 0 ,
\eqno (19)$$

$$
\int_V 
{\partial n  \over \partial x_j}
dv_1 dv_2 dv_3 = 0 ,
\eqno (20)$$

\par
This means, that averages of $n$ derivatives on space and time coordinates
are zero.

\par
Let us integrate (17) on velocities and use (20). We get identity:

$$
\int_V 
{\partial ( n v_k )  \over \partial x_k} 
dv_1 dv_2 dv_3 = 0 .
\eqno (21)$$

\par
But we know, that average velocity $u_k$ is equal to

$$
\rho u_k = \int_V 
n v_k
dv_1 dv_2 dv_3 .
\eqno (22)$$

\par
Therefore (22) is similar to (7) - the divergence of average velocity field is zero. 
We can write this in the following way:

$$
\int_V 
{\partial n  \over \partial x_k} v_k
dv_1 dv_2 dv_3 = 0 .
\eqno (23)$$

\par
This identity concerns the first moments of derivatives of $n$ on space coordinates. They are not zero, but their sum is zero.

\par
Let us multiply (17) by $v_j$ and integrate the result by parts:

$$
{\partial ( n v_j )  \over \partial t} +
{\partial ( n v_k v_j )  \over \partial x_k} +
{\partial ( n b_k v_j )  \over \partial v_k} =
n b_j .
\eqno (24)$$

\par
Let us integrate (25) on velocities. We get expression for average accelerations 
(compare this with (5))

$$
\rho ( b_j )_{avg} =
\int_V 
n b_j
dv_1 dv_2 dv_3 =
\int_V 
\left(
{\partial ( n v_j )  \over \partial t} +
{\partial ( n v_k v_j )  \over \partial x_k}
\right)\ 
dv_1 dv_2 dv_3
 .
\eqno (25)$$

\par
Let us denote

$$
J_{kj} =
\int_V n v_k v_j dv_1 dv_2 dv_3 = 
\rho u_k u_j - \sigma_{kj} ,
\eqno (26)$$

\par\noindent
where $\sigma_{kj}$ - components of stresses tensor.

\par
So (26) reads

$$
\rho ( b_j )_{avg} =
{\partial ( \rho u_j )  \over \partial t} +
{\partial ( \rho u_k u_j )  \over \partial x_k} -
{\partial \sigma_{kj}   \over \partial x_k} =
\rho a_j - {\partial \sigma_{kj}   \over \partial x_k}
 .
\eqno (27)$$

\par
This is very remarkable, that average acceleration is not equal to acceleration of average
motion and contain additional term - divergence of stresses tensor. This follows from the fact, that accelerations are quadratic on velocities.

\par
We shall use (27) to get constraint expression.

\par
Integrate Newton's law (18) and get:

$$
\rho ( b_j )_{avg} = 
\left(
Q_j
\right)_{avg} =
F_j .
\eqno (28)$$

\par
(27) and (28) imply equations of movement :

$$
\rho a_j - {\partial \sigma_{kj}   \over \partial x_k} = 
F_j
 .
\eqno (29)$$

\par
The Gauss functional without constraints has the following form

$$
G =
\int_X
dx_1 dx_2 dx_3 
\int_V 
{n \over 2 }\left(
b_k - {1 \over m }Q_k
\right)\ 
\left(
b_k - {1 \over m }Q_k
\right)\ 
dv_1 dv_2 dv_3 
 .
\eqno (30)$$

\par
To get expression for constraint on accelerations we take divergence of (27) (compare with (9)).

$$
\int_V 
{\partial   \over \partial x_j}
\left(
n b_j
\right)\ 
dv_1 dv_2 dv_3 =
\int_V 
{\partial   \over \partial x_j}
\left(
{\partial ( n v_j )  \over \partial t} +
{\partial ( n v_k v_j )  \over \partial x_k}
\right)\ 
dv_1 dv_2 dv_3
 .
\eqno (31)$$

\par
The first term in the RHS of (31) is zero according to (21). The rest terms give us the desired constraint expression:

$$
\int_V 
\left[
{\partial   \over \partial x_j}
\left(
n b_j
\right) - 
{\partial^2 n   \over \partial x_j \partial x_k} v_k v_j
\right]
dv_1 dv_2 dv_3 = 0 .
\eqno (32)$$

\par
We multiply (32) by $p / \rho$ and add the result to (31). Gauss functional with constraint reads now (compare with (10)):

$$
G =
\int_X
dx_1 dx_2 dx_3 
\int_V 
\left\{
{n \over 2 }\left(
b_k - {1 \over m }Q_k
\right)\ 
\left(
b_k - {1 \over m }Q_k
\right)\ 
-
\right.
\eqno (33)$$
$$
\left.
-
{p (x_i )  \over \rho }\left[
{\partial   \over \partial x_j}
\left(
n b_j
\right) - 
{\partial^2 n   \over \partial x_j \partial x_k} v_k v_j
\right]\ 
\right\}
dv_1 dv_2 dv_3 
 .
$$

\par
We seek the extreme of functional (33) with $(n b_j )$ as unknown variables.
The Euler's equation for (33) is:

$$
n b_k = 
{n \over m }Q_k - 
{1 \over \rho }n {\partial p  \over \partial x_k} .
\eqno (34)$$

\par
This means, that reaction of constrains is really equivalent to pressure field and we return to (12) and (14) once again.

\section{Fokker - Planck equation for incompressible fluid in curvilinear coordinates}

\par
Now let us consider the form of (13-14) system in the curvilinear coordinates. We need not perform all calculations here, because we can refer for details to our previous works [1-2]. Therefore we give here only results of calculations.

\par
Let us denote $g_{mn}$ - the covariant components of metric tensor and
$g^{mn}$ - the contravariant components of metric tensor, 
$g = det | g_{ij} | = 1 / det | g^{ij} |$ .

\par
Christoffel's symbol is called
$$
\Gamma_p ,_{mn} = {1 \over 2 }\left( 
{\partial g_{np}   \over \partial x^m } +
{\partial g_{pm}   \over \partial x^n } -
{\partial g_{mn}   \over \partial x^p }
\right) .
\eqno (35)$$

\par\noindent
and
$$
\Gamma_{mn}^p = g^{pq} \Gamma_q ,_{mn} .
\eqno (36)$$

\par
Covariant components of velocity vector are $v_i$, contravariant
components of velocity vector are $v^i$.

\par
Using these definitions, we can write  equations (13-14) in 
curvilinear coordinates with contravariant velocities 
as independent variables, as

$$
\int_V n\  
\sqrt g\  dv^1 dv^2 dv^3 = 1 .
\eqno (37)$$

$$
{\partial n  \over \partial t } +
v^k {\partial n  \over \partial x^k } -
\Gamma_{pq}^k
v^p v^q
{\partial n  \over \partial v^k } -
\alpha\  v^k
{\partial n  \over \partial v^k } - 
3\  \alpha\  n -
\eqno (38)$$
$$
-
{1 \over \rho }g^{mn}
{\partial n  \over \partial v^m}
{\partial p  \over \partial x^n}
=
k\  g^{lk}
{\partial^2 n  \over \partial v^l \partial v^k } .
$$

\par
For details we refer to our works [1-2].

\par
For example, the system (37-38) in spherical coordinates reads:

$$
\int_V n\  
dv^1 dv^2 dv^3 = 
{\rho \over r^2 \sin ( \theta )} .
\eqno (39)$$

$$
{\partial n  \over \partial t } +
v^1 {\partial n  \over \partial r } +
v^2 {\partial n  \over \partial \theta } +
v^3 {\partial n  \over \partial \phi } +
\eqno (40)$$
$$
+ r \left( v^2 v^2 + \sin^2 ( \theta ) v^3 v^3
\right){\partial n  \over \partial v^1 } +
\left( \sin ( \theta ) \cos ( \theta )v^3 v^3 - {2 \over r }v^1
v^2 \right) {\partial n  \over \partial v^2 } -
2 \left( v^1 v^3 + {\cos ( \theta )   \over \sin ( \theta ) } v^2 v^3 \right) {\partial n  \over \partial v^3 } -
$$
$$
-
\alpha\  
\left(
v^1 {\partial n  \over \partial v^1 }
+ v^2 {\partial n  \over \partial v^2 }
+ v^3 {\partial n  \over \partial v^3 }
\right)
- 3\  \alpha\  n -
{1 \over \rho }\left(
{\partial n  \over \partial v^1}
{\partial p  \over \partial r} +
{1 \over r^2}
{\partial n  \over \partial v^2}
{\partial p  \over \partial \theta} +
{1 \over r^2 \sin^2 ( \theta )}
{\partial n  \over \partial v^3}
{\partial p  \over \partial \phi}
\right)
=
$$
$$
=
k\  
\left(
{\partial^2 n  \over \partial v^1 \partial v^1 }
+ {1 \over r^2} {\partial^2 n  \over \partial v^2 \partial v^2 }
+ {1 \over r^2 \sin^2 ( \theta ) } {\partial^2 n  \over \partial v^3 \partial v^3 }
\right) .
$$

\par
The system (13-14) in curvilinear coordinates and with 
covariant velocities as independent variables, is

$$
\int_V n\  
{1 \over \sqrt g} dv_1 dv_2 dv_3 = \rho .
\eqno (41)$$

$$
{\partial n  \over \partial t } +
g^{mk} v_m {\partial n  \over \partial x^k } +
\Gamma_{kl}^q
g^{pl}
v_p v_q
{\partial n  \over \partial v_k } -
\alpha\  v_k 
{\partial n  \over \partial v_k } - 
3\  \alpha\  n -
\eqno (42)$$
$$
-
{1 \over \rho }{\partial n  \over \partial v_k}
{\partial p  \over \partial x^k}
=
k\  g_{lk}
{\partial^2 n  \over \partial v_l \partial v_k } .
$$

\par
For example, the system (41-42) in spherical coordinates has the form:

$$
\int_V n\  
dv_1 dv_2 dv_3 = r^2 \sin ( \theta ) \rho .
\eqno (43)$$

$$
{\partial n  \over \partial t } +
v_1 {\partial n  \over \partial r } +
{ v_2   \over r^2} {\partial n  \over \partial \theta } +
{ v_3   \over r^2 \sin^2 ( \theta ) } {\partial n  \over \partial \phi } +
\eqno (44)$$
$$
+ {1 \over r^3 }\left( 
v_2 v_2 + {v_3 v_3   \over \sin^2 ( \theta ) } 
\right) {\partial n  \over \partial v_1 } +
{\cos ( \theta ) v_3 v_3   \over r^2 \sin^3 ( \theta ) }
{\partial n  \over \partial v_2 } - 
$$
$$
-
 \alpha\  
\left(
v_1 {\partial n  \over \partial v_1 } 
+ v_2 {\partial n  \over \partial v_2 } 
+ v_3 {\partial n  \over \partial v_3 } 
\right)
- 3\  \alpha\  n -
{1 \over \rho }\left(
{\partial n  \over \partial v^1}
{\partial p  \over \partial r} +
{\partial n  \over \partial v^2}
{\partial p  \over \partial \theta} +
{\partial n  \over \partial v^3}
{\partial p  \over \partial \phi}
\right)
=
$$
$$
=
k\  
\left(
{\partial^2 n  \over \partial v_1 \partial v_1 } 
+ r^2 {\partial^2 n  \over \partial v_2 \partial v_2 } 
+ r^2 \sin^2 ( \theta ) {\partial^2 n  \over \partial v_3
\partial v_3 } 
\right) .
$$

\par
For orthogonal coordinates the diagonal components of metric tensor
are expressed as squares of $H_i$ - Lame coefficients. All
off-diagonal components are zero. The Christoffel's symbol components
can be expressed as derivatives of the Lame coefficients.

\par
The system (13-14) in curvilinear coordinates and with 
physical velocities as independent variables, is

$$
\int_V n\  
dw^1 dw^2 dw^3 = \rho .
\eqno (45)$$

$$
{\partial n  \over \partial t } +
{w^k  \over H_k}
{\partial n  \over \partial x^k } +
{\partial n  \over \partial w^k } 
{w^s  \over H_s H_k}
\left(
w^s {\partial H_s  \over \partial x^k } -
w^k {\partial H_k  \over \partial x^s} 
\right)
- \alpha\  w^k
{\partial n  \over \partial w^k }
- 3\  \alpha\  n -
\eqno (46)$$
$$
-
{1 \over \rho } {1 \over H_k}
{\partial n  \over \partial w^k}
{\partial p  \over \partial x^k}
=
k\  
{\partial^2 n  \over \partial w^i \partial w^i } . \ \  (s !=
k)
$$

\par
For example, the system (45-46) in spherical coordinates has the form:

$$
\int_V n\  
dw^1 dw^2 dw^3 = \rho .
\eqno (47)$$

$$
{\partial n  \over \partial t } +
{w^1} {\partial n  \over \partial r } +
{w^2  \over r }{\partial n  \over \partial \theta } +
{w^3  \over r \sin ( \theta ) }{\partial n  \over \partial \phi } +
\eqno (48)$$
$$
+ {1 \over r }\left( w^2 w^2 + w^3 w^3 \right) {\partial
n  \over \partial w^1 } +
{1 \over r }\left( {\cos ( \theta )   \over \sin ( \theta ) } w^3 w^3 - w^1 w^2 \right) {\partial n  \over \partial w^2 } -
{1 \over r }\left( w^1 w^3 + {\cos ( \theta )   \over \sin ( \theta
) } w^2 w^3 \right) {\partial n  \over \partial w^3 } -
$$
$$
-\  \alpha\  
\left(
w^1 {\partial n  \over \partial w^1 } +
w^2 {\partial n  \over \partial w^2 } +
w^3 {\partial n  \over \partial w^3 }
\right)
- 3\  \alpha\  n -
$$
$$
-
{1 \over \rho }\left(
{\partial n  \over \partial v^1}
{\partial p  \over \partial r} +
{1 \over r }{\partial n  \over \partial v^2}
{\partial p  \over \partial \theta} +
{1 \over r \sin ( \theta )}
{\partial n  \over \partial v^3}
{\partial p  \over \partial \phi}
\right)
=
k\  
\left(
{\partial^2 n  \over \partial w^1 \partial w^1 } +
{\partial^2 n  \over \partial w^2 \partial w^2 } +
{\partial^2 n  \over \partial w^3 \partial w^3 }
\right) .
$$

\section{Stationary solution of Fokker - Planck equation for incompressible fluid}

\par
Only in this section we discuss the case of nonzero force field. In this case
equation (14) reads

$$
{\partial n  \over \partial t} +
v_j {\partial n  \over \partial x_j} - 
\alpha\  {\partial \over \partial v_j} (v_j n) +
{1 \over \rho }{\partial n  \over \partial v_j}
\left(
F_j -
{\partial p  \over \partial x_j}
\right) = 
k\  {\partial^2 n  \over \partial v_j \partial v_j} .
\eqno (49)$$

\par\noindent
where $F_i ( \vec x ) $ are components of force acting
on particle. For non potential forces the stationary solution of Fokker - Planck equation does not exist. For potential forces we have the following expressions for force
components:

$$
F_i =
- {\partial Pi  \over \partial x_i} ,
\eqno (50)$$

\par\noindent
and (14) reads

$$
{\partial n  \over \partial t} +
v_j {\partial n  \over \partial x_j} - 
\alpha\  {\partial \over \partial v_j} (v_j n) -
{\partial n  \over \partial v_j}
{\partial   \over \partial x_j}
\left(
Pi + p
\right) = 
k\  {\partial^2 n  \over \partial v_j \partial v_j} ;
\eqno (51)$$

\par\noindent
where $Pi ( \vec x )$ -  potential function.

\par
Stationary solution of usual Fokker - Planck equation has the form
$n = m( \vec x )\ s( \vec v )$. For the case of incompressible fluid
must be $m( \vec x ) = 1$ because of condition (13). Therefore
$n = n( \vec v )$. 

\par
To kill all the terms  with cross products of derivatives on space coordinates
and velocities in (51), we substitute:

$$
p = const - Pi .
\eqno (52)$$

\par
This is the Pascal's law.

\par
The rest of terms in (51), depending only on velocities, is
the divergence of some current in velocities space.
For the true static solution all components of the current 
must be zero:

$$
- \alpha v_j n =
k {\partial n  \over \partial v_j} .
\eqno (53)$$

\par
So $n$ has Maxwell distribution :

$$
n = n_0 
\exp \left[
- {\alpha \over 2k }v_j v_j
\right] .
\eqno (54)$$

\par
The value of $n_0$ constant factor we find from condition (13)

$$
n_0 
= \rho\  \left( {\alpha \over 2 \pi k} \right)^{3/2} .
\eqno (55)$$

$$
n = \rho\  \left( {\alpha \over 2 \pi k} \right)^{3/2} 
\exp \left[
- {\alpha \over 2k }v_j v_j
\right] .
\eqno (56)$$

\section{Linearization of Fokker - Planck equation for incompressible fluid}

\par
We seek solution of (13-14) as a sum of stationary solution
and perturbation term:

$$
p = \epsilon \bar p .
\eqno (57)$$

$$
n = \rho \left( {\alpha \over 2 \pi k} \right)^{3/2} 
\exp \left[
- {\alpha \over 2k }v_j v_j
\right] +
\epsilon \bar n .
\eqno (58)$$

\par\noindent
($Pi = 0$ here).

\par
Substitute these expressions to (13-14) and drop terms
of more than first degree on $\epsilon$.

\boxit{
$$
\int_V \bar n dv_x dv_y dv_z = 0 .
\eqno (59)$$
}

\boxit{
$$
{\partial \bar n   \over \partial t} +
v_j {\partial \bar n   \over \partial x_j} - 
\alpha\  {\partial \over \partial v_j} (v_j \bar n ) +
\left(
{\alpha \over k }\right)\ 
\left( {\alpha \over 2 \pi k} \right)^{3/2} 
\exp \left[
- {\alpha \over 2k }v_j v_j
\right]\ 
v_k
{\partial \bar p   \over \partial x_k} 
= k\  {\partial^2 \bar n   \over \partial v_j \partial v_j} .
\eqno (60)$$
}

\par
The system (59-60) though linear, is not trivial. We shall discuss it's solution later.

\section{Symmetries of Fokker - Planck equation for incompressible fluid}

\par
We presented symmetries of standard Fokker - Planck equation (1) in our previous work [6] with all calculations details. In this section we present only result of calculations for Fokker - Planck equation for incompressible fluid.

\par
Two equations of the (13-14) have different nature: (13) is integral equation and (14) is 
differential. To obtain symmetries of this system we perform two-step process: 1) we determine symmetries of differential equation (14) and 2) we calculate actions of these symmetries operators upon (13) and drop all invalid operators, for which action does not vanish.

\par
We omit tedious details of step 1) and give below only resulting expressions. Expressions for variations of variables are:

$$
\delta x = r_2 z-r_3 y + 
3\  C_1 e^{ \alpha / 2 t} x + f_1 (t);
\eqno (61)$$
$$
\delta y = r_3 x-r_1 z + 
3\  C_1 e^{{\alpha} t / 2} y + f_2 (t);
$$
$$
\delta z = r_1 y-r_2 x + 
3\  C_1 e^{{\alpha} t / 2} z + f_3 (t) .
$$

\par\noindent
where $r_1 , r_2 , r_3 , C_1$ - arbitrary constant coefficients; $f_1 , f_2 , f_3$ - arbitrary functions of argument $t$.

$$
\delta u = {3 \over 2 }\alpha\  C_1 e^{{\alpha} t / 2} x + 
C_1 e^{{\alpha} t / 2} u + 
r_2 w - r_3 v + 
f_1 ' (t);
\eqno (62)$$
$$
\delta v = {3 \over 2 }\alpha\  C_1 e^{{\alpha} t / 2} y + 
C_1 e^{{\alpha} t / 2} v + 
r_3 u - r_1 w + 
f_2 ' (t);
$$
$$
\delta w = {3 \over 2 }\alpha\  C_1 e^{{\alpha} t / 2} z + 
C_1 e^{{\alpha} t / 2} w + 
r_1 v - r_2 u + 
f_3 ' (t);
$$

$$
\delta t = {4 \over \alpha }C_1 e^{{\alpha} t / 2} + C_2 .
\eqno (63)$$

$$
\delta n =
f_4 
( n e^{{3} \alpha t} )\  e^{ - 3 \alpha t } -
12 C_1 e^{{\alpha} t / 2} n .
\eqno (64)$$

$$
\delta p = 
( \alpha f_1 ' - f_1 '' ) x +
( \alpha f_2 ' - f_2 '' ) y +
( \alpha f_3 ' - f_3 '' ) z +
\eqno (65)$$
$$
+
{3 \over 8 }\alpha^2 C_1 e^{{\alpha} t / 2} (x^2 + y^2 + z^2 ) + 
2 C_1 e^{{\alpha} t / 2} p +
f_5 (t) .
$$

\par\noindent
where $C_2$ - one more arbitrary constant coefficient; $f_4 , f_5$ - two more arbitrary functions of argument $t$.

\par
These expressions lead to the following expressions for symmetries operators:

\par\noindent
- operator (rather exotic) associated with $C_1$:

$$
v_1 =
e^{{\alpha} t / 2}
\left[
-12 n {\partial   \over \partial n} +
\left(
{3 \over 8 }\alpha^2 (x^2 + y^2 + z^2 ) +
2 p
\right)
 {\partial   \over \partial p} +
{4 \over \alpha }{\partial   \over \partial t} 
+
\right.
\eqno (66)$$
$$
+ \left.
\left( {3 \over 2 }\alpha x + u \right) {\partial   \over \partial u} +
\left( {3 \over 2 }\alpha y + v \right) {\partial   \over \partial v} +
\left( {3 \over 2 }\alpha z + w \right) {\partial   \over \partial w} +
3 x {\partial   \over \partial x} +
3 y {\partial   \over \partial y} +
3 z {\partial   \over \partial z} 
\right] ;
$$

\par\noindent
- time shift operator associated with $C_2$:

$$
v_2 =
{\partial   \over \partial t} ;
\eqno (67)$$

\par\noindent
- three operators of shifts along $x , y, z$ axes associated with $f_1 , f_2 , f_3$ functions:

$$
v_3 =
( \alpha f_1 ' - f_1 '' ) x {\partial   \over \partial p} +
f_1 ' (t) {\partial   \over \partial u} +
f_1 (t) {\partial   \over \partial x} ;
\eqno (68)$$

$$
v_4 =
( \alpha f_2 ' - f_2 '' ) y {\partial   \over \partial p} +
f_2 ' (t) {\partial   \over \partial v} +
f_2 (t) {\partial   \over \partial y} ;
\eqno (69)$$

$$
v_5 =
( \alpha f_3 ' - f_3 '' ) z {\partial   \over \partial p} +
f_3 ' (t) {\partial   \over \partial w} +
f_3 (t) {\partial   \over \partial z} ;
\eqno (70)$$

\par\noindent
- one more exotic operator associated with $f_4$:

$$
v_6 =
f_4 
( n e^{{3} \alpha t} )\  e^{ - 3 \alpha t } {\partial   \over \partial n} ;
\eqno (71)$$

\par\noindent
- operator, which states, that we can freely choose arbitrary additive pressure at each moment of time, associated with $f_5$:

$$
v_7 =
f_5 (t) {\partial   \over \partial p} ;
\eqno (72)$$

\par\noindent
- three time independent rotations:

$$
v_8 =
 v {\partial   \over \partial w} - 
 w {\partial   \over \partial v} +
 y {\partial   \over \partial z} - 
 z {\partial   \over \partial y} ;
\eqno (73)$$

$$
v_9 =
 w {\partial   \over \partial u} - 
 u {\partial   \over \partial w} +
 z {\partial   \over \partial x} - 
 x {\partial   \over \partial z} ;
\eqno (74)$$

$$
v_{10} =
 u {\partial   \over \partial v} - 
 v {\partial   \over \partial u} +
 x {\partial   \over \partial y} -
 y {\partial   \over \partial x} .
\eqno (75)$$

\par
Now expression for variation of (13) is:

$$
\int_V 
\left[
\delta n + n\  \left(
{\partial \delta u  \over \partial u} +
{\partial \delta v  \over \partial v} +
{\partial \delta w  \over \partial w} 
\right)
\right]\ 
dv_1 dv_2 dv_3 = 0 ,
\eqno (76)$$

\par
Substitute (66-75) to (76) and get:

$$
\int_V 
\left[
f_4 
( n e^{{3} \alpha t} )\  e^{ - 3 \alpha t } -
12 C_1 e^{{\alpha} t / 2} n + n 3 C_1 e^{{\alpha} t / 2}
\right]\ 
dv_1 dv_2 dv_3 = 0 ,
\eqno (77)$$

\par
This means

$$
C_1 = 0 ;
\eqno (78)$$

\par\noindent
and

$$
f_4 = 0 .
\eqno (79)$$

\par
Therefore symmetries of Fokker - Planck equation for incompressible fluid build the subgroup of (66-75) with excluded exotic symmetries $v_1$ and $v_6$.

\par
This list of symmetries is the same as the Navier - Stokes equations symmetries list (see ref. [7]), with exception of scaling symmetries. Our system has no scaling symmetries, because it contain additive frictional term.

\section{Symmetries of linearized equations}

\par
Similar to previous section, we present only result of symmetries calculations without calculations details.

\par
We use the same two-step process. As a result of the first step we get following expressions for variations of variables:

$$
\delta n =
C_1 n +
\bar n ,
\eqno (80)$$

$$
\delta p = 
C_1 p +
\bar p ,
\eqno (81)$$

\par\noindent
where $C_1$ - arbitrary constant coefficient, $\bar n$, $\bar p$ - arbitrary solutions of linearized system (59-60) itself;

$$
\delta t = C_2 ,
\eqno (82)$$

\par\noindent
where $C_2$ - one more arbitrary constant coefficient;

$$
\delta x = r_2 z-r_3 y + 
C_3 ,
\eqno (83)$$
$$
\delta y = r_3 x-r_1 z + 
C_4 ,
$$
$$
\delta z = r_1 y-r_2 x + 
C_5 ,
$$

\par\noindent
where $r_1 , r_2 , r_3 , C_3 , C_4 , C_5$ - arbitrary constant coefficients;

$$
\delta u = r_2 w - r_3 v ,
\eqno (84)$$
$$
\delta v = r_3 u - r_1 w ,
$$
$$
\delta w = r_1 v - r_2 u .
$$

\par
These expressions lead to the following expressions for symmetries operators:

\par\noindent
- scaling operator associated with $C_1$:

$$
v_1 =
n {\partial   \over \partial n} + p {\partial   \over \partial p} ;
\eqno (85)$$

\par\noindent
- time shift operator associated with $C_2$:

$$
v_2 =
{\partial   \over \partial t} ;
\eqno (86)$$

\par\noindent
- three operators of shifts along $x , y, z$ axes associated with $C_3 , C_4 , C_5$ :

$$
v_3 =
{\partial   \over \partial x} ;
\eqno (87)$$

$$
v_4 =
{\partial   \over \partial y} ;
\eqno (88)$$

$$
v_5 =
{\partial   \over \partial z} ;
\eqno (89)$$

\par\noindent
- three rotations:

$$
v_6 =
 v {\partial   \over \partial w} - 
 w {\partial   \over \partial v} +
 y {\partial   \over \partial z} - 
 z {\partial   \over \partial y} ;
\eqno (90)$$

$$
v_7 =
 w {\partial   \over \partial u} - 
 u {\partial   \over \partial w} +
 z {\partial   \over \partial x} - 
 x {\partial   \over \partial z} ;
\eqno (91)$$

$$
v_8 =
 u {\partial   \over \partial v} - 
 v {\partial   \over \partial u} +
 x {\partial   \over \partial y} -
 y {\partial   \over \partial x} ;
\eqno (92)$$

\par\noindent
- infinite subgroup of linear system (59-60) solutions:

$$
v_9 =
 \bar n {\partial   \over \partial n} + 
 \bar p {\partial   \over \partial p} .
\eqno (93)$$

\par
Now we go to second step. Expression for variation of (59) is:

$$
\int_V 
\left[
\delta n + n\  \left(
{\partial \delta u  \over \partial u} +
{\partial \delta v  \over \partial v} +
{\partial \delta w  \over \partial w} 
\right)
\right]\ 
dv_1 dv_2 dv_3 = 0 ,
\eqno (94)$$

\par
Substitute (80-84) to (94) and get:

$$
\int_V 
(
C_1 n
)\ 
dv_1 dv_2 dv_3 = 0 ,
\eqno (95)$$

\par
This means

$$
C_1 = 0 .
\eqno (96)$$

\par
Therefore symmetries of linearized Fokker - Planck equation for incompressible fluid build the subgroup of (85-93) with excluded scaling symmetry $v_1$.

\par
Let us give some small examples of invariant solutions.

\par
Rotations (90-92) have following invariants:

$$
r = \sqrt { x^2 + y^2 + z^2 } ;
\eqno (97)$$

$$
U = \sqrt { u^2 + v^2 + w^2 } ;
\eqno (98)$$

$$
U_r =
{ x u + y v + z w } .
\eqno (99)$$

\par
According to PDE symmetries theory (ref. [7]), we should seek solutions of (59-60) of the following form:

$$
n = n ( t, r, U, U_r );
\eqno (100)$$

$$
p = p ( t, r);
\eqno (101)$$

\par
(2) reads now

$$
{\partial n   \over \partial t} +
{ U_r   \over r }{\partial n  \over \partial r} +
U^2 {\partial n  \over \partial U_r} -
\alpha U {\partial n  \over \partial U} -
\alpha U_r {\partial n  \over \partial U_r} -
3 \alpha n +
\eqno (102)$$
$$
+
\left(
{\alpha \over k }\right)\ 
\left( {\alpha \over 2 \pi k} \right)^{3/2} 
\exp \left[
- {\alpha \over 2k }U^2
\right]\ 
{ U_r   \over r }{\partial p  \over \partial r}
= k\  \left(
{\partial^2 n  \over \partial U^2} +
{2 \over U }{\partial n  \over \partial U} +
2 {U_r \over U }{\partial^2 n  \over \partial U_r \partial U} +
r^2 {\partial^2 n  \over \partial U_r^2}
\right) .
$$

\par
In this way we reduced (60) to equation with 4 independent variables. This equation is still not easy to solve. To give simpler example we drop $U_r$ and keep only $U$ variable:

$$
n = n ( t, r, U ).
\eqno (103)$$

\par
Then

$$
{\partial n   \over \partial t} +
{ U_r   \over r }{\partial n  \over \partial r} -
\alpha U {\partial n  \over \partial U} -
3 \alpha n +
\eqno (104)$$
$$
+
\left(
{\alpha \over k }\right)\ 
\left( {\alpha \over 2 \pi k} \right)^{3/2} 
\exp \left[
- {\alpha \over 2k }U^2
\right]\ 
{ U_r   \over r }{\partial p  \over \partial r}
= k\  \left(
{\partial^2 n  \over \partial U^2} +
{2 \over U }{\partial n  \over \partial U}
\right) .
$$

\par
Equate coefficient by $U_r$ to zero

$$
{\partial n  \over \partial r} +
\left(
{\alpha \over k }\right)\ 
\left( {\alpha \over 2 \pi k} \right)^{3/2} 
\exp \left[
- {\alpha \over 2k }U^2
\right]\ 
{\partial p  \over \partial r}
= 0 .
\eqno (105)$$

\par
The rest of (104) is

$$
{\partial n   \over \partial t} -
\alpha U {\partial n  \over \partial U} -
3 \alpha n = k\  \left(
{\partial^2 n  \over \partial U^2} +
{2 \over U }{\partial n  \over \partial U}
\right) .
\eqno (106)$$

\par
Integration of (106) gives

$$
n +
\left(
{\alpha \over k }\right)\ 
\left( {\alpha \over 2 \pi k} \right)^{3/2} 
\exp \left[
- {\alpha \over 2k }U^2
\right]\ 
p
= f(t, U) .
\eqno (107)$$

\par
Apply (59) to (107). We see, that $p$ depends only on $t$

$$
p = g(t) .
\eqno (108)$$

\par
This means, that ${\partial n  \over \partial r} = 0$ also and we need only solve (29) for $n (t, U)$.

\shead{DISCUSSION}
\par\noindent
In this article we deducted the Fokker - Planck equation for incompressible fluid from the Gauss principle. This method can be useful for investigation of another cases of constrained movement. To promote this task we give three slightly different forms of equations for use of curvilinear coordinates. 
\par\noindent
Unfortunately, the actual solution of equations is a complicated task, because we move from the model with totally independent particles to model with interaction. This model is by necessity nonlinear. To make first steps to solution we use perturbation method. We construct linearized equations, but do not try to solve them at the moment - this is a task for another work.
\par\noindent
It is of some interest to obtain symmetries group for our system. We find the group and compare it with symmetries of Navier - Stokes equations. One symmetry is missing in our case as a result of friction force presence.
\par\noindent
We derive symmetries of linearized equations also and give some examples of invariant solutions. We see, that symmetries group of linearized system is not rich enough and is hardly usable to obtain physically interesting solutions.

\shead{ACKNOWLEDGMENTS}
\par\noindent
We wish to thank Jos A. M. Vermaseren from NIKHEF (the Dutch Institute for Nuclear 
and High-Energy Physics), for he made his symbolic computations program FORM
available for download for non-commercial purposes (see [8]).

\rule{2in}{1pt}
\shead{REFERENCES}

\begin{IPlist}
\IPitem{{[1]}}
Igor A. Tanski. Fokker - Planck equation in curvilinear coordinates. 

arXiv:nlin.CD/0506053 v1 25 Jun 2005

\IPitem{{[2]}}
Igor A. Tanski. Fokker - Planck equation in curvilinear coordinates. 

Part 2. arXiv:nlin.CD/0512024 v1 10 Dec 2005

\IPitem{{[3]}}
Whittaker E. T., A Treatise on the Analytical Dynamics
of Particles and Rigid Bodies, third edition.
Cambridge University press, 1927.

\IPitem{{[4]}}
Zak M. A. Neklassicheskie problemy mehaniki sploshnyh sred. Lenigrad university press, 1974.

\IPitem{{[5]}}
Frank A. M. Diskretnye modeli neszhimaemoj zhidkosti. FIZMATLIT, 2001. ISBN 5-9221-0190-0

\IPitem{{[6]}}
Igor A. Tanski. The symmetries of the Fokker - Planck equation in three dimensions. v1 nlin.CD/0501017 8 Jan 2005

\IPitem{{[7]}}
L. V. Ovsiannikov, Group Analysis of Differential Equations, English translation edited by W. F. Ames (Academic, New York, 1982).

\IPitem{{[8]}}
Michael M. Tung.
FORM Matters: Fast Symbolic Computation under UNIX.

arXiv:cs.SC/0409048 v1 27 Sep 2004

\end{IPlist}

\end{document}